\documentclass[aps,preprint,epsfig,rotate]{revtex4}
\begin{document}
\title{Free gravitational field in the metric gravity as a Hamilton system}
 \author{Alexei M. Frolov}
 \email[E--mail address: ]{afrolov@uwo.ca}

\affiliation{Department of Applied Mathematics \\
 University of Western Ontario, London, Ontario N6H 5B7, Canada}

\date{\today}

\begin{abstract}

The closed system of Hamilton equations is derived for all tensor components of the free gravitational field  $g_{\alpha\beta}$ and corresponding momenta $\pi^{\gamma\delta}$
in the metric General Relativity. The Hamilton-Jacobi equation for the free gravitational field $g_{\alpha\beta}$ is also derived and discussed. In general, all methods and 
procedures based on the Hamilton and Hamilton-Jacobi approaches are very effective in actual applications to many problems known in the metric GR. 
 
\noindent 
PACS number(s): 04.20.Fy and 11.10.Ef

\end{abstract}

\maketitle
\newpage

\section{Introduction}

In this communication we derive the both Hamiltonian and Hamiltonian-Jacobi equations for the free gravitational field in the metric General Relativity (or GR, for short). 
In our earlier studies \cite{K&K} and \cite{FK&K} we have developed the two non-contradictory Hamilton approaches for the metric GR. Also, in \cite{K&K} and \cite{FK&K} 
we derived the explicit formulas for the total $H_t$ and canonical $H_C$ Hamiltonians in the both approaches. Furthermore, it was shown in \cite{FK&K} that these two 
Hamiltonian-based formulations of metric GR are related to each other by some canonical transformation of all dynamical variables, i.e., by transformations of the 
corresponding `momenta' and `coordinates'. Such canonical transformations are allowed transformations in any Hamiltonian-based approach and they are often used to simplify 
analysis of various Hamiltonian systems, including dynamical systems combined from different fields and particles. Theory developed below is substantially based on the 
procedure proposed by Dirac \cite{Dir50} - \cite{Dir58} and results obtained in our earlier studies \cite{K&K} and \cite{FK&K} with the use of this Dirac's procedure. A 
number of fundamental facts known for the general Hamilton-Jacobi theory \cite{Gant} - \cite{Arnold} are extensively used below. Since the volume of this contribution is 
limited, below we restrict ourselves to the derivation of basic equations only.      

First of all, we need to introduce a few principal notations. Everywhere in this study the notations $g_{\alpha\beta}$ stand for the covariant components of the metric 
tensor which are dimensionless functions. Analogous notations $\pi^{\alpha\beta}$ designate the corresponding contravariant components of momenta which conjugate to the 
$g_{\alpha\beta}$ components (for more detail, see \cite{K&K} and \cite{FK&K}). The determinant of the metric $g_{\alpha\beta}$ tensor is denoted by its traditional 
notation $- g (> 0)$. The Latin alphabet is used for spatial components, while the index 0 means the temporal component (or time-component). In this study the notation 
$d$ designates the total dimension of the space-time manifold, where $d \ge 3$ \cite{X}. This means that an arbitrary Greek index $\alpha$ varies between 0 and $d - 1$, 
while an arbitrary Latin index varies between 1 and $d - 1$. The quantaties and tensors such as $B^{((\alpha \beta) \gamma | \mu \nu \lambda)}, I_{mnpq}$, etc, have been 
defined in earlier papers \cite{K&K}, \cite{FK&K} and \cite{Dir58}. In this study the definitions of all these quantaties and tensors are exactly the same as in \cite{K&K} 
and \cite{FK&K} and here we do not want to repeat them. The short notations $g_{\alpha\beta,k}$ and $g_{\gamma\rho,0}$ are used below for the spatial and temporal 
derivatives of the corresponding components of the metric tensor. Any expression which contains a pair of identical (or repeated) indexes, where one index is covariant 
and another is contravariant, means summation over this `dummy' index. This convention is very convenient and drastically simplifies the explicit formulas derived in the 
metric GR.   

Now, we can introduce the Lagrangian of the metric general relativity. The original Einstein-Hilbert Lagrangian (see, e.g., \cite{Carm}), which was used in the first 
papers on the metric GR, cannot be applied for the purposes of this study, since it contains the second-order derivative(s). However, the same Einstein-Hilbert Lagrangian 
can be transformed into the $\Gamma - \Gamma$ Lagrangian which is reduced to the following general form
\begin{eqnarray}
  {\cal L} &=& \frac14 \sqrt{-g} B^{\alpha\beta\gamma\mu\nu\rho} \Bigl(\frac{\partial g_{\alpha\beta}}{\partial x^{\gamma}}\Bigr) \Bigl(\frac{\partial 
  g_{\mu\nu}}{\partial x^{\rho}}\Bigr) = \frac14 \sqrt{-g} B^{\alpha\beta\gamma\mu\nu\rho} g_{\alpha\beta,\gamma} g_{\mu\nu,\rho} \label{eq05} 
\end{eqnarray}
where $B^{\alpha\beta\gamma\mu\nu\rho} = g^{\alpha\beta} g^{\gamma\rho} g^{\mu\nu} - g^{\alpha\mu} g^{\beta\nu} g^{\gamma\rho} + 2 g^{\alpha\rho} g^{\beta\nu} g^{\gamma\mu} 
- 2 g^{\alpha\beta} g^{\gamma\mu} g^{\nu\rho}$ is a homogeneous cubic function of the contravariant components of the metric tensor $g^{\alpha\beta}$. In the right-hand side 
of this formula, Eq.(\ref{eq05}), we have used the short notation $g_{\alpha\beta,\gamma}$ to designate the partial derivatives $\frac{\partial g_{\alpha\beta}}{\partial 
x^{\gamma}}$. Note that the $\Gamma - \Gamma$ Lagrangian ${\cal L}$, Eq.(\ref{eq05}), contains the partial temporal derivatives $g_{0\sigma,0}$ of the first-order only, and 
it is used below to derive the actual Hamiltonian of metric GR. In fact, to derive the closed formula for the Hamiltonian of metric GR we need a slightly different form of 
the $\Gamma - \Gamma$ Lagrangian which explicitly contains all temporal derivatives (or time-derivatives) is (see, e.g., \cite{K&K})  
\begin{eqnarray}
  {\cal L} = \frac14 \sqrt{-g} B^{\alpha\beta 0\mu\nu 0} g_{\alpha\beta,0} g_{\mu\nu,0} + \frac12 \sqrt{-g} B^{(\alpha\beta 0|\mu\nu k)} g_{\alpha\beta,0} g_{\mu\nu,k} 
   + \frac14 \sqrt{-g} B^{\alpha\beta k \mu\nu l} g_{\alpha\beta,k} g_{\mu\nu,l} \label{eq51}
\end{eqnarray}
where the notation $B^{(\alpha\beta\gamma|\mu\nu\rho)}$ means a `symmetrical' $B^{\alpha\beta\gamma\mu\nu\rho}$ quantity which is symmetrized in respect to the permutation 
of the two groups of indexes, i.e.,
\begin{eqnarray}
   B^{(\alpha\beta\gamma|\mu\nu\rho)} = \frac12 \Bigl( B^{\alpha\beta\gamma\mu\nu\rho} + B^{\mu\nu\rho\alpha\beta\gamma} \Bigr) \label{eq52}
\end{eqnarray}

By using the Lagrangian ${\cal L}$, Eq.(\ref{eq51}), and standard definition of momenta (see, e.g., \cite{Dir64}), we obtain the explicit formulas for all components of the 
momenta tensor $\pi^{\gamma\sigma}$   
\begin{eqnarray}
  \pi^{\gamma\sigma} = \frac{\partial {\cal L}}{\partial g_{\gamma\sigma,0}} = \frac{1}{2} \sqrt{-g} B^{((\gamma\sigma) 0|\mu\nu 0)} g_{\mu\nu, 0} 
   + \frac{1}{2} \sqrt{-g} B^{((\gamma\sigma) 0|\mu\nu k)} g_{\mu\nu, k} \; \; \; \label{mom}
\end{eqnarray}
The first term in the right-hand side of the last equation can be written in the form 
\begin{eqnarray}
 \frac{1}{2} \sqrt{-g} B^{((\gamma\sigma)0|\mu\nu 0)} g_{\mu\nu, 0} = \frac{1}{2} \sqrt{-g} g^{00} E^{\mu\nu\gamma\sigma} g_{\mu\nu, 0} \; \; 
 \label{B}
\end{eqnarray}
where the space-like tensor (also called the Dirac tensor) $E^{\mu\nu\gamma\sigma}$ is
\begin{eqnarray}
  E^{\mu \nu \gamma \rho} = e^{\mu \nu} e^{\gamma \rho} - e^{\mu \gamma} e^{\nu \rho} \; \; , \; \; {\rm where} \; \; \; e^{\mu \nu} = g^{\mu \nu} -
  \frac{g^{0 \mu} g^{0 \nu}}{g^{00}} \; \; \; \label{E}  
\end{eqnarray}
As follows form the formula, Eq.(\ref{E}), the tensor $e^{\mu \nu}$ equals zero, if either index $\mu$, or index $\nu$ (or both) equals zero. The same statement is true for 
the $E^{\mu\nu\gamma\sigma}$ tensor. It is also clear that the space-like tensor $E^{p q k l}$ is not equal zero and this tensor is always invertable. Below, we shall 
designate the corresponding inverse (space-like) tensor as $I_{m n q p}$, i.e., we can write $I_{m n q p} E^{p q k l} = \delta^{k}_{m} \delta^{l}_{n}$.

First, let us consider the `regular' case when $\gamma = p$ and $\sigma = q$. In this case one finds
\begin{eqnarray}
  \pi^{pq} = \frac{\partial {\cal L}}{\partial g_{p q,0}} = \frac{1}{2} \sqrt{-g} B^{((p q) 0|\mu\nu 0)} g_{\mu\nu,0} \label{momenta}
   + \frac{1}{2} \sqrt{-g} B^{((p q) 0|\mu\nu k)} g_{\mu\nu, k}
\end{eqnarray}
This equation is invertable and the velocity $g_{m n, 0}$ is expressed as the linear function (or linear combination) of the space-like components $\pi^{pq}$ of 
the momentum tensor:
\begin{eqnarray}
  g_{mn, 0} = \frac{1}{g^{00}} \Bigl( \frac{2}{\sqrt{-g}} I_{m n q p} \pi^{pq} - I_{m n q p} B^{((pq) 0|\mu\nu k)} g_{\mu\nu, k} \Bigr) =
  \frac{1}{g^{00}} I_{m n q p} \Bigl( \frac{2}{\sqrt{-g}} \pi^{pq} - B^{((pq) 0|\mu\nu k)} g_{\mu\nu, k} \Bigr) \label{veloc}
\end{eqnarray}
Formally, for the space-like components of metric tensor $g_{pq}$ and corresponding momenta $\pi^{mn}$ one finds no principal differences with the dynamical systems, 
which are routinely studied in classical mechanics. Indeed, the Hamiltonian  

However, as follows from Eq.(\ref{veloc}) in the metric gravity the corresponsing velocities $g_{pq,0}$ and momenta 
$\pi^{pq}$ are not related by one scalar parameter (some `effective' mass). Instead, for the free gravitational field(s) such a relation, Eq.(\ref{veloc}), has a matrix 
form and one velocity linearly depends upon a number of different space-like components of momenta $\pi^{mn}$ and vice versa. 

In the second `non-regular' (or singular) case, when $\gamma = 0$, the first term in the righ-hand side of Eq.(\ref{mom}) equals zero and this equation takes the 
from 
\begin{eqnarray}
  \pi^{0\sigma} = \frac{\partial {\cal L}}{\partial g_{0\sigma,0}} = \frac{1}{2} \sqrt{-g} B^{((0\sigma) 0|\mu\nu k)} g_{\mu\nu, k} 
   \; \; \; \label{constr}
\end{eqnarray}
Note that this equation does not contain any velocity et al. Furthermore, this equation, Eq.(\ref{constr}), determines the momentum $\pi^{0\sigma}$ as an analytical 
(qubic) functions of the contravariant components of the metric tensor $g^{\alpha\beta}$ and spatial derivatives of its covariant components $g_{\mu\nu, k}$. It is 
clear that such a situation cannot be found neither in classical, nor in quantum mechanics of arbitrary systems of particles. However, for various physical fields 
similar situations arise quite often. The physical meaning of Eq.(\ref{constr}) is simple and can be expressed in the following words. The function
\begin{eqnarray}
   \phi^{0\sigma} = \pi^{0\sigma} - \frac{1}{2} \sqrt{-g} B^{((0\sigma) 0|\mu\nu k)} g_{\mu\nu, k} 
\end{eqnarray}
must be equal zero at any time, i.e., it does not change during the actual physical motion (or evolution) of the actual gravitational field. Dirac in \cite{Dir50} 
proposed to write such equalities in the symbolic form $\phi^{0\sigma} \approx 0$, where the $d-$functions $\phi^{0\sigma}$ (for $\sigma = 0, 1, \ldots, d - 1$) are 
called the primary constraints.  

\section{Hamiltonian(s) of the metric General Relativity}

The knowledge of the Lagrangian ${\cal L}$, represented above in the form, Eq.(\ref{eq51}), allows us to derive the closed formula for the Hamiltonian of the metric GR. 
The total Hamiltonian $H_t$ of the free gravitational field in metric GR derived in \cite{K&K} is written in our notations in the form 
\begin{eqnarray}
  H_t = \pi^{\alpha\beta} g_{\alpha\beta,0} - {\cal L} = H_C + g_{0\sigma,0} \phi^{0\sigma}  \label{eq1}
\end{eqnarray}
where ${\cal L}$ is the $\Gamma - \Gamma$ Lagrangian of the metric GR, $\phi^{0\sigma}$ are the primary constraints $\phi^{0\sigma} = \pi^{0\sigma} - \frac{1}{2}\sqrt{-g} 
B^{\left( \left(0\sigma\right) 0\mid\mu\nu k\right)} g_{\mu\nu,k}$, while $g_{0\sigma,0}$ are the corresponding $\sigma-$velocities' and $H_C$ is the canonical 
Hamiltonian of the metric GR
\begin{eqnarray}
 & &H_C = \frac{1}{\sqrt{-g} g^{00}} I_{mnpq} \pi^{mn} \pi^{pq} - \frac{1}{g^{00}} I_{mnpq} \pi^{mn} B^{(p q 0|\mu \nu k)} g_{\mu\nu,k} \label{eq5} \\
 &+& \frac14 \sqrt{-g} \Bigl[ \frac{1}{g^{00}} I_{mnpq} B^{((mn)0|\mu\nu k)} B^{(pq0|\alpha\beta l)} - B^{\mu\nu k \alpha\beta l}\Bigr] g_{\mu\nu,k} g_{\alpha\beta,l} 
 \nonumber
\end{eqnarray}
which does not contain any primary constraint $\phi^{0\sigma}$. The primary constraints arise during transition from the $\Gamma - \Gamma$ Lagrangian ${\cal L}$ to the 
Hamiltonians $H_t$ and $H_C$, since the $\Gamma - \Gamma$ Lagrangian ${\cal L}$ is a linear (not quadratic!) function of all $d$ momenta $\pi^{0\sigma} = \frac{\delta 
L}{\delta g_{0\sigma,0}}$ each of which include at least one temporal index \cite{K&K}. The Poisson brackets (see, Eqs.(\ref{eq15}) and (\ref{grPB}) below) between all 
primary constraints equal zero identically, i.e., $[ \phi^{0\sigma}, \phi^{0\gamma}] = 0$. This allows us to predict that the Poisson brackets between the primary 
constraints $\phi^{0\gamma}$ and Hamiltonian $H_C$ cannot be proportional to any of the primary constraints. Therefore, these Poisson brackets $[\phi^{0\sigma}, H_C]$ 
lead one directly to the secondary constraints $\chi^{0\sigma}$, i.e., $[ \phi^{0\sigma}, H_C] = \chi^{0\sigma}$. The explicit formulas for the secondary constraints
$\chi^{0\sigma}$ are \cite{K&K}
\begin{eqnarray}
 \chi^{0\sigma} &=& -\frac{g^{0\sigma}}{2 \sqrt{-g} g^{00}} I_{mnpq} \pi^{mn} \pi^{pq} + \frac{g^{0\sigma}}{2 g^{00}} I_{mnpq} \pi^{mn} 
  A^{( pq0 \mid \mu\nu k )} g_{\mu\nu,k} + \delta_{m}^{\sigma} \Bigl[ \pi_{,k}^{mk} + \Bigl(\pi^{pk} e^{qm}  \label{eqn8} \\
 &-& \frac12 \pi^{pq} e^{km}) g_{pq,k} \Bigr] - \frac{\sqrt{-g}}{8} \Bigl(\frac{g^{0\sigma}}{g^{00}} I_{mnpq} B^{((mn) 0 \mid \mu \nu k)} B^{(pq0 \mid \alpha \beta t)} 
 - g^{0\sigma} B^{\mu \nu k \alpha \beta t} \Bigr) g_{\mu\nu,k} g_{\alpha\beta,t} \nonumber \\
 &+& \frac{\sqrt{-g}}{4 g^{00}} I_{mnpq} B^{((mn) 0 \mid\mu\nu k )} g_{\mu\nu,k} g_{\alpha\beta,t} 
 \Bigl[ g^{\sigma t} \Bigl( g^{00} g^{p \alpha} g^{q \beta} + g^{pq} g^{0 \alpha} g^{0 \beta} - 2 g^{\alpha q} g^{0 p} g^{0 \beta} \Bigr) \nonumber \\
 &-& g^{\sigma p} \Bigl( 2 g^{00} g^{q \alpha} g^{t \beta} - g^{00} g^{\alpha \beta} g^{q t} + g^{\alpha\beta} g^{0q} g^{0t} - 2 g^{q \alpha} g^{0 \beta} g^{0t} 
 - 2 g^{t \alpha} g^{0 \beta} g^{0q} + 2 g^{qt} g^{0\alpha} g^{0\beta} \Bigr) \nonumber \\ 
 &+& g^{0\sigma} ( 2 g^{\beta t} g^{\alpha p} g^{0q} - 2 g^{p\alpha} g^{q\beta} g^{0t} - 2 g^{pq} g^{t\beta} g^{0\alpha} + 2 g^{pt} g^{q\beta} g^{0\alpha} + 
 g^{pq} g^{\alpha\beta} g^{0t} - g^{tp} g^{\alpha\beta} g^{0q}) \Bigr] \nonumber \\
 &-& \frac{\sqrt{-g}}{4} g_{\mu\nu,k} g_{\alpha\beta,t} \Bigl[ g^{\sigma t} ( g^{\alpha\mu} g^{\beta\nu} g^{0k} + g^{\mu\nu} g^{\alpha t} g^{0\beta} - 2 g^{\mu\alpha} 
 g^{k\nu} g^{0\beta} ) \nonumber \\
 &+& g^{0\sigma} ( 2 g^{\alpha t} g^{\beta\mu} g^{\nu k} - 3 g^{t\mu} g^{\nu k} g^{\alpha\beta} - 2 g^{\mu\alpha} g^{\nu\beta} g^{kt} + g^{\mu\nu} g^{kt}
g^{\alpha\beta} + 2 g^{\mu t} g^{\nu\beta} g^{k\alpha}) \nonumber \\
 &+& g^{\sigma\mu} \Bigl( (g^{\alpha\beta} g^{\nu t} - 2 g^{\nu\alpha} g^{t\beta}) g^{0k} + 2 ( g^{\beta\nu} g^{kt} - g^{\beta k} g^{t\nu}) g^{0\alpha} + 
 ( 2 g^{k\beta} g^{\alpha t} - g^{\alpha\beta} g^{kt}) g^{0\nu}\Bigr) \Bigr] \nonumber \\
&-& \frac{\sqrt{-g} g^{00}}{2} E^{pqt\sigma} \Bigl( \frac{1}{g^{00}} I_{mnpq} B^{((mn)0 \mid \mu\nu k)} g_{\mu\nu,k} \Bigr)_{,t}
  - \frac{\sqrt{-g}}{2} B^{((\sigma 0) k \mid \alpha \beta t)} g_{\alpha\beta,kt} \nonumber
\end{eqnarray}
where 
\begin{eqnarray}
 A^{\alpha\beta 0 \mu\nu k}= B^{(\alpha\beta 0 \mid \mu \nu k)} - g^{0k} E^{\alpha \beta \mu \nu} + 2 g^{0\mu} E^{\alpha \beta k \nu}.
\end{eqnarray}
and $\sigma = 0, 1, \ldots, d - 1$. This means that we have $d-$secondary constraints, i.e., the total number of all found primary and secondary constraints for 
this Hamilton formulation equals $d + d = 2 d$. Note also that all these primary and secondary constraints $\phi^{0\gamma}$ and $\chi^{0\sigma}$, where $\sigma = 
0, 1, \ldots, d - 1$, are the first-class constraints \cite{Dir64}. The Poisson brackets between $\phi^{0\gamma}$ and $\chi^{0\sigma}$ constraints, which are of 
interest in this study, is given by the following relations $[ \phi^{0\gamma}, \phi^{0\mu} ] = 0$ and $[ \phi^{0\gamma}, \chi^{0\sigma}] = \frac12 g^{\gamma\sigma}$.  
The Poisson brackets between canonical Hamiltonian $H_C$ and secondary constraints $\chi^{0\sigma}$ are expressed as `quasi-linear' \cite{QL} combinations of the 
same secondary constrains $\chi^{0\sigma}$, i.e., 
\begin{eqnarray}
 [ \chi^{0\sigma}, H_{c} ] &=& -\frac{2}{\sqrt{-g}} I_{mnpq} \pi^{mn} \frac{g^{\sigma q}}{g^{00}} \chi^{0p} + \frac12 g^{\sigma k} g_{00,k} \chi^{00} + 
 \delta_{0}^{\sigma} \chi_{,k}^{0k} \label{eqn122} \\
 &+& \Bigl( -2 \frac{1}{\sqrt{-g}} I_{mnpk} \pi^{mn} \frac{g^{\sigma p}}{g^{00}} + I_{mkpq} g_{\alpha\beta,t} \frac{g^{\sigma m}}{g^{00}} A^{(pq) 0 \alpha\beta t} 
 \Bigr)\chi^{0k} \nonumber \\
 &-& \Bigl( g^{0\sigma} g_{00,k} + 2 g^{n\sigma} g_{0n,k} + \frac{g^{n\sigma} g^{0m}}{g^{00}} (g_{mn,k} + g_{km,n} - g_{kn,m}) \Bigr) \chi^{0k} \nonumber
\end{eqnarray}
This relation indicates that the Hamilton procedure developed for the metric GR in \cite{K&K} and \cite{FK&K} is closed, i.e., it does not lead to any tertiary, or 
other constraints of higher order(s). Briefly, this proves that the found system of the canonical Hamiltonian $H_C$ and all primary and secondary constraints is closed 
in terms of some simplectic geometry which is determined by the Poisson brackets (their definition is presented below). In other words, we proved the complete closure 
of the Dirac procedure \cite{K&K} for the metric GR which is described by the canonical $H_C$ Hamiltonian, Eq.(\ref{eq5}).  

Note that the formulas for the Hamiltonians $H_t, H_C$ presented above and explicit expressions for all secondary constraints \cite{K&K} allow one to derive (with the 
use of Castellani procedure \cite{Cast}) the correct generators of gauge transformations, which directly lead to the diffeomorphism invariance, i.e., to the invariance 
which is well known for the free gravitational field(s) in the metric GR (see, e.g., \cite{Carm}, more details and discussion can be found in \cite{K&K}). Such a
reconstruction of the diffeomorphism invariance is a relatively simple problem for the Lagrangian-based approaches which are often used in the metric GR (see, e.g., 
\cite{Saman}). In contrast with this, for any Hamiltonian-based approach the complete solution of similar problem requires a substantial work. On the other hand, 
analytical and explicit derivation of the diffeomorphism invariance becomes a very good test for the total $H_t$ and canonical $H_C$ Hamiltonians and for all primary 
$\phi^{0\sigma}$ and secondary $\chi^{0\sigma}$ constraints derived in any new Hamiltonian approach which is proposed for the metric GR. To this moment only two 
Hamiltonian-based approaches developed in \cite{K&K} and \cite{FK&K} (see also \cite{Dir58}) were able to reproduce the complete diffeomorphism invariance. On the other 
hand, there are many `advanced' Hamiltonian-based procedures known in the metric GR which cannot pass this fundamental test, in principle. In particular, the long list 
of such `unlucky' Hamilton-like procedures includes the ADM gravity which is often called the `geometro-dynamics'. The basic principles of ADM gravity were developed 
and discussed in \cite{ADM} and \cite{Regg}. The modern status of this theory is well described in \cite{Wald1} and \cite{Wald2}. The fact that the four-dimensional 
diffeomorphism "was somehow lost in the chain of analytical transformations" \cite{Ish} in the Hamiltonian formulation of the ADM gravity is known since the middle of 
1960's (see, e.g., discussion in \cite{Ish} and references therein). Analysis of the fundamental reasons of such a failure of ADM formulation can be found in \cite{KKN}.   

Now, we need to define the Poisson brackets between two arbitrary (analytical) functions of the $\pi^{\alpha\beta}$ and $g_{\rho\gamma}$ variables. In general, the properly 
defined Poisson brackets play a central role in the construction of an arbitrary Hamilton (dynamical) system. Note, that analytical calculation of the Poisson brackets 
between two arbitrary functions of dynamical variables in Hamilton GR is always reduced to the calculation of the fundamental Poisson brackets which are defined as follows
\begin{eqnarray}
  [ g_{\alpha\beta}, \pi^{\mu\nu}] = - [ \pi^{\mu\nu}, g_{\alpha\beta}] = g_{\alpha\beta} \pi^{\mu\nu} - \pi^{\mu\nu} g_{\alpha\beta} = \frac12 
  \Bigl(\delta^{\mu}_{\alpha} \delta^{\nu}_{\beta} + \delta^{\nu}_{\alpha} \delta^{\mu}_{\beta}\Bigr) = \Delta^{\mu\nu}_{\alpha\beta} \; \; \; ,  \label{eq15} 
\end{eqnarray}
where the symbol $\delta^{\mu}_{\beta}$ is the Kronecker delta, while the notation $\Delta^{\mu\nu}_{\alpha\beta}$ stands for the gravitational (or tensor) delta-function.
All other fundamental Poisson brackets between dynamical variables of the metric GR equal zero identically, i.e., $[ g_{\alpha\beta}, g_{\mu\nu}] = 0$ and 
$[ \pi^{\alpha\beta}, \pi^{\mu\nu}] = 0$. This indicates clearly that the dynamical variables $g_{\alpha\beta}$ and $\pi^{\mu\nu}$ (see, \cite{K&K}, \cite{FK&K}) in the 
metric GR have been defined correctly, and now we can apply these variables and fundamental Poisson brackets to analyze and solve various gravitational problems. For 
instance, based on Eq.(\ref{eq15}) we can define the gravitational Poisson bracket of the two arbitrary functions $U$ and $V$ each of which depends upon the dynamical 
variables $g_{\alpha\beta}$ and $\pi^{\mu\nu}$. This definition is
\begin{eqnarray}
  [ U, V ] = \sum_{(\alpha\beta)} \Bigl( \frac{\partial U}{\partial g_{\alpha\beta}} \frac{\partial V}{\partial \pi^{\alpha\beta}} - \frac{\partial U}{\partial 
 \pi^{\alpha\beta}} \frac{\partial V}{\partial g_{\alpha\beta}} \Bigr) \; \; \; .  \label{grPB}
\end{eqnarray}
It is easy to check that these Poisson brackets obey all general rules and conditions known for the Poisson brackets (see, e.g., \S 21 in \cite{Dir79}). For instance, for 
the Poisson bracket between a regular function $F(g_{\alpha\beta})$, which depends upon the components of metric tensor $g_{\alpha\beta}$ only, and $\gamma\rho$-component 
of the momentum tensor ($\pi^{\gamma\rho}$) one finds
\begin{eqnarray}
  [ F(g_{\alpha\beta}), \pi^{\gamma\rho} ] = \frac{\partial F}{\partial g_{\alpha\beta}} \Delta^{\gamma\rho}_{\alpha\beta} = \frac{\partial F}{\partial g_{\gamma\rho}} 
  \; , \; \; \label{eq254}
\end{eqnarray}
This formula can easily be obtained, e.g., by representing the function $F(g_{\alpha\beta})$ by its Maclaurin and/or Taylor series. 

\section{Hamilton equations for the free gravitational field} 

By using the formulas, Eqs.(\ref{eq1}) and (\ref{eq5}), for the total and canonical Hamiltonians, i.e., for $H_t$ and $H_C$, respectively, we obtain the following 
Hamilton equations (or system of Hamilton equations) which describe time-evolution of all dynamical variables in the metric GR, i.e., time-evolution of each component 
of the $g_{\alpha\beta}$ and $\pi^{\gamma\rho}$ tensors. These equations are
\begin{eqnarray}
 \frac{d g_{\alpha\beta}}{d x_0} = [ g_{\alpha\beta}, H_{t} ] \; \; \; {\rm and} \; \; \;  \frac{d \pi^{\gamma\rho}}{d x_0} = [ \pi^{\gamma\rho}, H_{t} ] \label{eq20}
\end{eqnarray}
where the notation $x_0$ denotes the temporal variable. In particular, for the spatial components $g_{ij}$ of the metric tensor one finds the following equations
\begin{eqnarray}
 \frac{d g_{ij}}{d x_0} = [ g_{ij}, H_{c} ] = \frac{2}{\sqrt{-g} g^{00}} I_{((ij)pq)} \pi^{pq} - \frac{1}{g^{00}} I_{(ij)pq} B^{(p q 0|\mu \nu k)} g_{\mu\nu,k}
 \label{eq25}
\end{eqnarray}
where the notations $I_{(ij)pq}$ and $I_{((ij)pq)}$ stand for the following `symmetrized' values
\begin{eqnarray}
 I_{(ij)pq} = \frac12 \Bigl( I_{ijpq} + I_{jipq} \Bigr) \; \; , \; and \; \; \; I_{((ij)pq)} = \frac12 \Bigl( I_{(ij)pq} + I_{(ij)qp} \Bigr) \label{eq251}
\end{eqnarray}
Analogously, for the $g_{0\sigma}$ components of the metric tensor one finds the following equations of time-evolution
\begin{eqnarray}
 \frac{d g_{0\sigma}}{d x_0} = [ g_{o\sigma}, H_{t} ] = g_{0\sigma,0} \; , \; \; \label{eq253}
\end{eqnarray}
since all $g_{0\sigma}$ components commute with the canonical Hamiltonian $H_C$, Eq.(\ref{eq5}). This result could be expected, since the equation, Eq.(\ref{eq253}), 
is, in fact, a definition of the $\sigma-$velocities, or the $g_{0\sigma,0}$-values, where $\sigma = 0, 1, \ldots, d - 1$.   

The Hamilton equations for the tensor components of momenta $\pi^{\alpha\beta}$, Eq.(\ref{eq20}), are substantially more complicated. They are derived by calculating 
the Poisson brackets between $H_{t}$ and $\pi^{\gamma\rho}$. This leads us to the following general formula
\begin{eqnarray}
 \frac{d \pi^{\gamma\rho}}{d x_0} &=& - [ H_{t}, \pi^{\gamma\rho} ] = - \Bigl[ \frac{\partial}{\partial g_{\gamma\rho}} \Bigl(\frac{I_{mnpq}}{\sqrt{-g} 
 g^{00}}\Bigr)\Bigr] \pi^{mn} \pi^{pq} + \Bigl[ \frac{\partial}{\partial g_{\gamma\rho}} \Bigl(\frac{I_{mnpq}}{g^{00}}\Bigr)\Bigr] \pi^{mn} 
 B^{(p q 0|\mu \nu k)} g_{\mu\nu,k} \nonumber \\
 &+&  \frac{1}{g^{00}} I_{mnpq} \pi^{mn}\Bigl[ \frac{\partial}{\partial g_{\gamma\rho}} B^{(p q 0|\mu \nu k)} \Bigr] g_{\mu\nu,k} + \frac12 
 \Delta^{\gamma\rho}_{0\sigma} \Bigl[ \frac{\partial}{\partial g_{0\sigma}} \Bigl(\sqrt{-g} B^{((0 \sigma) 0|\mu \nu k)}\Bigr) \Bigr] g_{\mu\nu,k} \label{eq255} \\
 &-& \frac14 \frac{\partial}{\partial g_{\gamma\rho}} \Bigl[ \frac{\sqrt{-g}}{g^{00}} I_{mnpq} B^{((mn)0|\mu\nu k)} B^{(pq0|\alpha\beta l)} - \sqrt{-g} 
 B^{\mu\nu k \alpha\beta l} \Bigr] g_{\mu\nu,k} g_{\alpha\beta,l} \nonumber
\end{eqnarray}

The arising Hamilton equations for the dynamical variables of metric GR, i.e., for the components of the $g_{\alpha\beta}$ and $\pi^{\gamma\rho}$ tensors, are significantly 
more complicated than analogous equations for the electro-magnetic fields known in Maxwell's electrodynamics. Nevertheless, the Hamilton equations (see, Eqs.(\ref{eq25}), 
(\ref{eq253}) and (\ref{eq255}) above), which govern the complete time-evolution of the free gravitational field(s) in the metric GR, have been derived in this study, and 
now one can apply them to describe these fields. Note that these equations, i.e., Eq.(\ref{eq25}), Eq.(\ref{eq253}) and Eq.(\ref{eq255}), form a conservative (or autonomous) 
system of differential equations. Theory of such equations and systems of such equations is a well developed area of modern Mathematics (see, e.g., \cite{Cart}, \cite{Arnold2} 
and references therein). For instance, we can say that every phase trajectory of this system of differential equations belongs to one of three following types: (a) a smooth 
curve without self-intersections, (b) a closed smooth curve (cycle), and (c) a point. There are many other facts known from the general theory of conservative systems of 
differential equations, and all of them can be used to determine the explicit form of the $g_{\alpha\beta}(x_0)$ and $\pi^{\gamma\rho}(x_0)$ dependencies. Furthermore, the 
arising (Hamilton) system of differential equations (see, Eqs.(\ref{eq25}), (\ref{eq253}) and Eq.(\ref{eq255}) above) conserves the total phase volume, i.e., the free 
gravitational field is a Liouville's dynamical system. In general, for any conservative (Hamilton) system of differential equations one finds a number of useful theorems and 
each of these theorems can now be used to solve (or simplify) the actual problems existing in the modern metric GR.

In general, any direct solution of the Hamilton equations for the free gravitational field(s) is a complex process, since these equations are essentially non-linear upon the
both covariant and/or contravariant components of the metric tensor. However, there are other methods which are often used to obtain different solutions of the Hamilton 
equations. All these methods are based on the Jacobi equation which is considered below. These methods are very powerful, since they allow one to detrmine not only a single 
solution of the canonical Hamilton equations, but also to obtain the partial and even complete integrals of motion for the free $d-$dimensional gravitational field(s). In 
fact, these integrals of motion form the closed Lie algebras with relatively simple (algebraic) commutation relations. Therefore, in actual applications one needs to determine 
only a few integrals of motion, while other similar integrals can be found by using these integrals and commutation relations (or Poisson brackets) between them.

\section{General Methodology of the Hamilton procedures}

Let us discuss the general methodology of Hamilton procedure which explains a number of advantages which this approach provides in applications to many problems, including 
metric General Relativity. 
In general, the Hamilton mechanics can be defined as a geometry in the phase space (see, e.g., \cite{Gant} and \cite{Arnold}). This means that by determining some dynamical (or 
mechanical) systems as a Hamilton system we reduce the original (e.g., mechanical) problem to some geometrical problem which is usually formulated in very clear and transparent 
mathematical (even geometrical) form. However, the main achievement of the Hamiltonian approach is an introduction of the double geometric structure, or double metric(s) in the 
phase space. One of these metrics is the fundamental metric (e.g., Euclidean, pseudo-Euclidean, Riemannian, etc) which is determined by the scalar products between two arbitrary 
vectors in the original multi-dimensional space. The second (or additional) `symplectic' metric is uniformly defined by the Hamiltonian system itself (or Poisson brackets). Note 
that these two metrics are considered in the same space. A possibility to apply analytical transformations in the both metrics instantaneously provides a number of great 
advantages for obtaining analytical and numerical solutions of the Hamilton equations of motions. Such a situation can be compared, e.g., with the Lagrange method(s) (or methods 
based on the use of Lagrangians), where only one usual metrics in the configuration space is used. 
  
Below, we restrict ourselves to the analysis of our Hamilton approach developed above (see, also \cite{K&K} and \cite{FK&K}). As is shown above in the case of $d-$dimensional 
space-time our Hamiltonian-based approach developed above for the metric General Relativity (or GR) leads to the actual Hamilton theory with $d-$primary and $d-$secondary 
constraints. Thus, the total number of constraints in the metric GR equals $2 d$. The total dimension of the corresponding phase (tensor) space, which is often called the 
cotangent space \cite{Arnold} (or $\Bigl\{ g_{\alpha\beta}, \pi^{\gamma\rho} \Bigr\}-$space), is $N_t = d (d + 1)$, which is always even. As is well known (see, e.g., 
\cite{Arnold} and references therein), an arbitrary Hamiltonian system is defined by an even-dimensional manifold (or phase space), a symplectic structure defined on this 
manifold (which is uniformly related with the relative Poincar\'{e} integral invariant) and a scalar function on it (the Hamilton function, or Hamiltonian, for short). In other 
words, the Hamilton system in the metric GR is defined by the $N_t$-dimensional cotangent $\Bigl\{ g_{\alpha\beta}, \pi^{\gamma\rho} \Bigr\}-$space, the Hamiltonian $H_t$ and 
the corresponding symplectic structure which is uniformly determined by the two following differential $\omega^{2}$ and $\omega^{1}$ forms: 
\begin{eqnarray}
  \omega^{2} = d\pi^{\alpha\beta} \wedge dg_{\alpha\beta} \; \; \; , \; \; \; \omega^{1} = \pi^{\alpha\beta} dg_{\alpha\beta} \; \; \; \label{omega-2}
\end{eqnarray}
The one-dimensional integral, which contains the differential form $\omega^{1}$, is called the relative integral invariant of the metric GR. In some papers this value is also 
called the symplectic potential (also pre-symplectic potential). The relation between these two forms is simple and transparent: $\omega^{2} = d\omega^{1}$. 

In a large number of textbooks on classical mechanics the symplectic structure on the even-dimensional phase space is determined by the corresponding Poisson brackets between 
all basic dynamical variables, i.e., between the `coordinates' and `momenta'. In other words, now the Hamiltonian systems is defined by: (1) the even-dimensional manifold (or 
phase space), (2) the scalar function $H$ (or Hamiltonian) which is determined on this manifold and has continuous derivatives of all orders, and (3) the symplectic structure 
on this manifold (or phase space) which is determined by the corresponding (canonical) Poisson brackets. In the case of metric GR we have the properly defined Hamiltonian 
$H_t$, Poisson brackets between all covariant components of the metric tensor $g_{\alpha\beta}$ and contravariant components of the momenta $\pi^{\alpha\beta}$. This Hamilton 
system is considered in the $N_t$-dimensional (or $d ( d + 1)$-dimensional) cotangent $\Bigl\{ g_{\alpha\beta}, \pi^{\gamma\rho} \Bigr\}-$space which is even-dimensional space. 
Note that each of the differential $\omega^{2}$ and $\omega^{1}$ forms uniformly determines the correct and complete set of Poisson brackets (and vice-versa) which transforms 
the $N_t$-dimensional cotangent $\Bigl\{ g_{\alpha\beta}, \pi^{\gamma\rho} \Bigr\}-$space into a symplectic even-dimensional manifold.     

The definitions of the Hamilon dynamical systems presented and discuss above are absolutely correct only for non-constraint dynamical systems. For dynamical systems with actual 
system of the first-class constraints such a definition is not complete. Indeed, for such systems one finds another Hamilton system (or sub-system) which has the Hamiltonian 
$H_C$ (or the canonical Hamiltonian, see above). In the case of metric GR such a system is defined in the $N_C =  N_t - 2 d = d (d - 1)$-dimensional cotangent $\Bigl\{ g_{kl}, 
\pi^{mn} \Bigr\}-$space. The corresponding symplectic structure in this space (or manifold) is determined by the following differential $\omega^{2}_{C}$ and $\omega^{1}_{C}$ 
forms
\begin{eqnarray}
  \omega^{2}_{C} = d\pi^{mn} \wedge dg_{mn} \; \; \; , \; \; \; \omega^{1}_{C} = \pi^{kl} dg_{kl} \; \; \; \label{omegac-2}
\end{eqnarray}
where the two differential forms are also related to each other by the equation $\omega^{2}_C = d\omega^{1}_C$. Each of these two differential forms determines the correct 
and complete set of Poisson brackets in the even-dimensional $R^{N_C}$ space. Thus, in the case of constrained dynamical system we have two different Hamilton systems 
defined in the $N_t$- and $N_C$-dimensional symplectic spaces, respectively. Since these two symplectic spaces have different (but even!) dimensions (equal $d (d + 1)$ and 
$d (d - 1)$, respectively), they cannot be isomorph to each other, in principle. It is also clear that these two dynamical Hamilton systems must be considered indepently and 
simultaneously.

Note again that the both Hamilton systems, which are generated by the differential $\omega^{2}$ and $\omega^{2}_C$ forms can be correct, if (and only if) the both $R^{N_t}$ 
and $R^{N_C}$ spaces are even-dimensional. In other words, the correct Hamilton approach cannot be developed in those cases, when, e.g., the dimension of the $R^{N_C}$ space 
is odd. This means that for any Hamilton (dynamical) system the total number of actual first-class constraints must be even. This statement is true for the Maxwell 
electrodynamics, where one finds one primary and one secondary constraints (see, e.g., \cite{Dir64}, \cite{Fro2014}). In this case the corresponding dimensions are: $N_t = 8$ 
and $N_C = 6$ \cite{AB}. For the $d-$dimensional metric General Relativity we have $d-$primary and $d-$secondary constraints (i.e., $2 d$ first-class constraints total, see, 
e.g., \cite{K&K} and \cite{FK&K}). Therefore, we have $N_t = d (d + 1)$ and $N_C = N_t - 2 d = d (d - 1)$. If the metric GR is considered in our regular four-dimensional 
space-time, then one finds four primary and four secondary constraints (or eight essential constraints total). In this case, the $R^{N_t}$-space has dimension twenty, while 
and corresponding dimension of the $R^{N_C}$-space equals twelve. The systems of equations arising in the both $R^{20}-$ and $R^{12}-$spaces describe the two different 
Hamilton systems, which, however, are closely related to each other.   

\section{Jacobi equation for the free gravitational field}

It was shown in \cite{FK&K} that the two different Hamiltonian approaches, developed in \cite{K&K} and  \cite{Dir58} (see also \cite{FK&K}) are related by a simple canonical 
transformation of dynamical variables. Therefore, it is possible to discuss the generating functions for these canonical transformations and apply a number of powerful tools 
developed for such functions in the general Hamilton-Jacobi procedure. This fact also allows us to derive the Hamilton-Jacobi equation for the free gravitational field. Based 
on this equation we can develop an alternative method which can be used to solve (both analytically and numerically) some important problems which are currently known in the 
metric GR. Note that in many books and textbooks all methods based on the Hamilton-Jacobi equation are recognized as the most effective, transparent and powerful tool for 
theoretical analysis of various Hamilton systems and applications to such systems (see, e.g., \cite{Arnold} and \cite{GF}). Cornelius Lanczos in Chapter VIII of his famous 
book \cite{Lanc} compared `the partial differential equation of Hamilton-Jacobi' with the mountain top of the variational classical mechanics. Furthermore, he chose the 
following phrase from the Bible as an epigraph to this chapter (Exodus, chapter 3, verse 5): ``Put off thy shoes from off thy feet, for the place whereon thou standest is 
holy ground''. As follows from a number of arguments presented below such an opinion about Jacobi (or Hamilton-Jacobi) equation is absolutely correct and justifiable.     

The closed system of Hamilton equations for the metric GR derived above allows one to obtain the Jacobi equation. This can be achieved in a number of different ways. Probably, 
the simplest approach is to investigate the closed system of Hamilton equations under some different angle which is directly related to the general theory of partial differential 
equations (see, e.g., \cite{Tric}, \cite{Fedor}). First, we note that each of the Hamilton equations is the first-order differential equation upon the corresponding 
time-derivatives $\frac{d g_{\alpha\beta}}{d x_0}$ and $\frac{d \pi^{\rho\sigma}}{d x_0}$. This means that these Hamilton equations can be considered as a system of 
characteristic equations for some non-linear equation. As was shown in a large number of classical books (see, e.g., \cite{Tric}, \cite{Fedor} and \cite{Cour}) this non-linear 
equation is the Jacobi equation, which is often called the Hamilton-Jacobi equation. The direct physical sense of the Hamilton-Jacobi equation and its solutions is straitforward: 
{\it the extremal will remain extremal as it propagates, if it satisfies the Hamilton-Jacobi equation}. In other words, the Hamilton-Jacobi eqation is an important necessary 
condition for some currently known extremal to propagate in the nearest future as the `new' extremal. For the non-constrained dynamical 
systems the Hamilton-Jacobi equation takes the form 
\begin{eqnarray}
  \frac{\partial S}{\partial t} + H(t, y_1, \ldots, y_{n};  p_1, \ldots, p_{n}) = \frac{\partial S}{\partial t} + H(t, y_1, \ldots, y_{n}; \frac{\partial S}{\partial y_1}, 
  \ldots, \frac{\partial S}{\partial y_n}\Bigr) = 0 \; \; , \; \; \label{Jacobi}
\end{eqnarray}
where $H$ is the Hamiltonian of this dynamical systems, $t$ is the time (or temporal coordinate), while $y_1, \ldots, y_{n}$ are the dynamical coordinates of the system 
and $p_1 = \frac{\partial S}{\partial y_1}, \ldots, p_{n} = \frac{\partial S}{\partial y_n}$ are the corresponding momenta. It is clear that the Hamilton-Jacobi equation 
is only the necessary condition of the fact that the extremal will propagate as an extremal into the new infinitesimally extended area. The corresponding sufficient 
conditions are more complicated and they will be discussed elsewhere (see,e.g., \cite{GF}). 

The function $S(t, y_1, \ldots, y_{n})$ is called the Jacobi function. Below, we shall follow the general theory described in \cite{Fedor}. This theory can be applied 
successfully, if the two following conditions are obeyed: (a) the function $H({\bf x}, {\bf p})$ (or Hamiltonian), has all derivatives of the first and second order which 
are the continuous functions everywhere in the $R^{2 n}_{x,p}$-space; and (b) all derivatives $\frac{\partial H}{\partial p_1}, \ldots, \frac{\partial H}{\partial p_n}$ 
do not equal zero at the same time. Note that the arising equation, Eq.(\ref{Jacobi}), is the non-linear (quadratic) equation upon the $\frac{\partial S}{\partial y_k}$ 
derivatives, where $k = 1, \ldots, n$. However, this equation is always a linear equation upon the time-derivative (or temporal-derivative) of the Jacobi function 
$\frac{\partial S}{\partial t}$. 

For dynamical systems with first-class constraints the actual situation is more complicated. However, all arising problems can be solved, and finally, we arrive to the 
following Jacobi (or Hamilton-Jacobi) equation for the free gravitational field
\begin{eqnarray}
 &-& \Bigl(\frac{\partial S}{\partial x_0}\Bigr) = \frac{1}{\sqrt{-g} g^{00}} I_{mnpq} \Bigl(\frac{\partial S}{\partial g_{mn}}\Bigr) 
 \Bigl(\frac{\partial S}{\partial g_{pq}}\Bigr) - \frac{1}{g^{00}} I_{mnpq} \Bigl(\frac{\partial S}{\partial g_{mn}}\Bigr) B^{(p q 0|\mu \nu k)} g_{\mu\nu,k} \label{eq555} \\
 &+& \frac14 \sqrt{-g} \Bigl[ \frac{1}{g^{00}} I_{mnpq} B^{((mn)0|\mu\nu k)} B^{(pq0|\alpha\beta l)} - B^{\mu\nu k \alpha\beta l}\Bigr] g_{\mu\nu,k} g_{\alpha\beta,l}  
 \nonumber \\
 &+& g_{0\sigma,0} \Bigl[ \Bigl(\frac{\partial S}{\partial g_{0\sigma}}\Bigr) - \frac{1}{2}\sqrt{-g} B^{\left( \left(0\sigma\right) 0\mid\mu\nu k\right)} g_{\mu\nu,k} \Bigr] 
 \nonumber
\end{eqnarray}
where $S(x_0, \{ g_{mn} \})$ is the Jacobi function of the gravitation field(s), while $\frac{\partial S}{\partial x_0}$ and $\frac{\partial S}{\partial g_{mn}}$ are its 
corresponding derivatives. 

The explicit derivation of the Jacobi equation, Eq.(\ref{eq555}), for the free gravitational field is the main resuts of this study. In general, analytical solution of the 
Jacobi equation is a well developed procedure which provides many advantages in comparison to the direct solution of the Hamilton equations \cite{XX}. Moreover, the Jacobi 
equation, Eq.(\ref{eq555}), opens a new avenue for derivation of a number of true and/or adiabatic invariants for the free gravitational field. As was mentioned above in 
analytical mechanics (see, e.g., \cite{Arnold}) the methods based on the Hamilton-Jacobi equation are often called and considered as the most effective procedures ever 
created for solving the equations of motion known for an arbitrary, in principle, Hamilton system. In particular, all methods based on the Hamilton-Jacobi equation are 
usually very effective for dynamical systems with Hamiltonians which contain only a few relatively small powers of all essential momenta. This includes the free gravitational 
field in the metric gravity, where the corresponding Hamiltonian is a quadratic functions of space-like momenta $\pi^{mn}$, Eq.(\ref{eq5}). Furthermore, based on the 
Hamilton-Jacobi equation one can determine a number of `integrals of motion', derive the complete and closed algebras of such integrals and investigate the properties of 
such algebras and commutation relations between these integrals. The regular procedure, which allows one to obtain all integrals of motion for any Hamilton system (and even for
arbitrary physical system with the known Lagrangian), is based on the Noeter theorem(s) \cite{Noeth} (see, also \cite{GF}). For instance, by using this theorem of Noeter in 
1921 Bessel-Hagen \cite{Bessl} determined the correct and complete Lie algebra (nowadays this algebra is called the $SO(4,2)-$algebra \cite{Bar}) of fifteen `conservation laws' 
(or generators) which correspond to the Maxwell equations in vacuum. The modern theory of group representations solved the same problem only 45 years later (see discussion and 
references in \cite{Bar}).  
 
Another interesting reason to deal with the Hamilton-Jacobi equation is a possibility to apply a wide class of analytical transformations which reduce this equation to relatively 
simple forms. For instance, by introducing the new `absolute' time $t$, which is simply related with the temporal coordinate $x_0$ used in Eq.(\ref{eq555}) above: $d x_{0} = 
\sqrt{- g} g^{00} dt$, one can reduce the original Hamilton-Jacobi equation to the form 
\begin{eqnarray}
 &-& \Bigl(\frac{\partial S}{\partial t}\Bigr) = I_{mnpq} \Bigl(\frac{\partial S}{\partial g_{mn}}\Bigr) \Bigl(\frac{\partial S}{\partial g_{pq}}\Bigr) - \sqrt{- g} I_{mnpq} 
 \Bigl(\frac{\partial S}{\partial g_{mn}}\Bigr) B^{(p q 0|\mu \nu k)} g_{\mu\nu,k} \label{eq5553} \\
 &+& \frac14 (- g) \Bigl[ I_{mnpq} B^{((mn)0|\mu\nu k)} B^{(pq0|\alpha\beta l)} - g^{00} B^{\mu\nu k \alpha\beta l} \Bigr] g_{\mu\nu,k} 
 g_{\alpha\beta,l} \nonumber 
%
\end{eqnarray}
where we chose the gauge in which all $\sigma-$velocities (or $g_{0\sigma,0}-$factors) equal zero, i.e., $g_{0\sigma,0} = 0$ for $\sigma = 0, 1, \ldots, d - 1$. This form of the 
Jacobi equation can be useful to simplify and solve a number of gravitational problems. The Jacobi function $S$ in Eq.(\ref{eq5553}) depends upon the temporal coordinate $x_0$ 
(or $t$) and upon $n_{C} = \frac{N_C}{2} = \frac{d (d - 1)}{2}$ real functions each of which is the covariant space-space (covariant) component $g_{mn}$ of the metric tensor. 
To describe the stationary situations, when the gravitational field does not change with time, we can introduce the `short' Jacobi function $S_0(\{ g_{mn} \})$ which does not 
depend upon temporal coordinate. The relation between the complete ($S$) and short ($S_0$) Jacobi functions is simple: $S(\{ g_{\alpha\beta} \}) = S_0(\{ g_{mn} \}) - E t$ (here 
the notations from Eq.(\ref{eq5553}) are applied), where $E$ is the numerical (or scalar) parameter which is called the total energy of the system. There is no need to detect an 
aditional physical sense in our definition of the total energy, since any Hamilton formulation of the metric GR is essentially based on the coordinate structure of the phase 
space, i.e., the procedure is `frustratingly non-invariant' from the very beginning \cite{Arnold}. Now, from Eq.(\ref{eq5553}) one finds the following expression for the total 
energy $E$ of the free gravitational field    
\begin{eqnarray}
 &E& = I_{mnpq} \Bigl(\frac{\partial S_{0}}{\partial g_{mn}}\Bigr) \Bigl(\frac{\partial S_{0}}{\partial g_{pq}}\Bigr) - \sqrt{- g} I_{mnpq} \Bigl(\frac{\partial S_0}{\partial 
 g_{mn}}\Bigr) B^{(p q 0|\mu \nu k)} g_{\mu\nu,k} \label{eq5554} \\
 &+& \frac14 (- g) \Bigl[ I_{mnpq} B^{((mn)0|\mu\nu k)} B^{(pq0|\alpha\beta l)} - g^{00} B^{\mu\nu k \alpha\beta l} \Bigr] g_{\mu\nu,k} g_{\alpha\beta,l} \nonumber 
\end{eqnarray}
where $I_{mnpq}$ is the following space-like tensor
\begin{eqnarray}
 I_{mnpq} = \frac{1}{d-2} g_{mn} g_{pq} - g_{mp} g_{nq} = \frac{1}{d-2} \Bigl[  g_{mn} g_{pq} - (d - 2) g_{mp} g_{nq} \Bigr] \label{tensorI}
\end{eqnarray}
This space-like tensor is positively defined, i.e., all its eigenvalues are positive. Furthermore, the $I_{mnpq}$ tensor is inverse to another space-like tensor $E^{pqkl}$, 
which was introduced by Dirac in 1950's (see, Eq.(\ref{E}) above and discussion in \cite{K&K}). 

\section{Conclusion}

In this study we have developed the complete Hamilton approach to describe free gravitational field(s) in the metric General Relativity. Such a description is essentially based 
on the same methods which are routinely used for various Hamilton systems in different branches of physics. In particular, in our Hamilton approach we also need to consider and 
solve the system of Hamilton equations which arises for the free gravitational field in the metric GR. Another great advantage of this study is the explicit derivation of the 
Jacobi (or Hamilton-Jacobi) equation for the free gravitational field. All these equations have never been produced in earlier studies. Very likely, the methods based on the 
Jacobi equation will soon be recognized as the most simple and transparent methods ever developed to determine the actual gravitational field(s) in metric GR. Furthermore, these 
methods can provide many advantages in future theoretical investigations. In particular, it is clear that the Hamilton-Jacobi equation for the free gravitational fields is much 
easier to solve than to deal with the original Einstein equation. Indeed, by using our Hamilton-Jacobi equation(s) for the metric GR we can always apply a very powerful and 
reliable apparatus developed for Hamilton dynamical systems. In contrast with this, direct solutions of the Einstein equation(s) are usually based on a semi-empirical choice of 
some components of the metric tensor. At the next stage of the direct solution one needs to fit parameters in these components in order to satisfy the Einstein equation for all 
components of the metric tensor. Quite often this is not possible, and, therefore, it is necessary to apply some different choice of analytical expression(s) for the components 
of metric tensor. This means that the whole procedure must be repeated from the very beginning. Hopefully, soon the methods based on the both Hamilton and Hamilton-Jacobi 
equations will be transformed (after some additional development) into a number of working procedures which can be applied to various problems currently known in the metric GR. 

In conclusion, we want to note that since the end of 1930's it was widely assumed that it is impossible to define any non-trivial symplectic structure in the $R^{d(d+1)}-$space 
based on the semi-quadratic (upon velocities) Lagrangians similar to the $\Gamma - \Gamma$ Lagrangian, Eqs.(\ref{eq05}) and (\ref{eq51}) which is used in the metric GR (see, e.g., 
\cite{K&K}, \cite{FK&K}). However, in 1950 Dirac found \cite{Dir50} an elegant way which allows one to restore such a symplectic structure in the tensor $R^{d(d+1)}-$space, which 
is the regular, working space for the $d-$dimensional metric GR. His approach also allowes us to determine, in principle, all gauge transformations for any given semi-quadratic 
Lagrangian. Explicitly this procedure was developed by Castellani in \cite{Cast}. The following creation of the complete Hamilton approach for the metric gravity was just a 
`matter of technique'. The first truly correct Hamilton approach for the metric GR was developed and published by Dirac in 1958 \cite{Dir58}. In this approach Dirac introduced 
some analytical expressions which were called the primary, secondary, etc, constraints. The primary constraints are always closely (even linearly) related to the canonical momenta 
$\pi^{0\mu}$. 

In general, it can be shown that all correct Hamiltonians arising in the metric GR are the quadratic functions of the space-like components of momenta $\pi^{mn}$ \cite{K&K}, 
\cite{FK&K} and \cite{Dir58}. Moreover, any Hamilton method which correctly describes the free gravitational field in the metric GR must generate the $d-$primary and $d-$secondary 
constraints. Furhtermore, the arising $H_C$ and $H_t$ Hamiltonians must uniformly be related by a reversible canonical tranformation with the $H_C$ and $H_t$ Hamiltonians derived 
in \cite{K&K} and with the analogous Hamiltonians obtained by Dirac in \cite{Dir58} (see, also \cite{FK&K}). An additional requirement to any correct system of Hamiltonian 
equations in the metric GR is obvious: the system of Hamiltonian(s) and all $d-$primary and $d-$secondary constraints must be able to reproduce the complete diffeomorphism 
invariance which is the actual invariance for the free gravitational field in the metric GR. If some Hamilton approach obeys all these requirements, then it can be considered as 
another correct Hamilton procedure developed for the metric GR. Furthermore, such a procedure must be canonically related with the Hamilton procedures developed in \cite{K&K}, 
\cite{FK&K} and \cite{Dir58}.
       
Based on the results of a number of recent studies we can conclude that the original Dirac approach \cite{Dir58} and other similar methods \cite{K&K}, \cite{FK&K}, which are 
canonically related with Dirac procedure \cite{Dir58}, work very well for the metric GR. Note also that for more than 60 years which passed after creation of the original Hamilton 
procedure in 1958 \cite{Dir58} nobody could detect even a single internal contradiction in this Dirac approach. In general, any correct non-contradictory Hamilton approach (see 
above) developed for the metric gravity can be considered as the first step to perform quantization of the free gravitational field(s). In this form the problem was originally 
formulated by Dirac in 1950 \cite{Dir50} (see also discussion in \cite{Dir64}). The recent development of the Hamiltonian-based approaches, including our derivation of the Jacobi 
equation, is of great interest for future theoretical research in the metric GR. 
    
I am grateful to my friends N. Kiriushcheva, S.V. Kuzmin and D.G.C. (Gerry) McKeon (all from the University of Western Ontario, London, Ontario, Canada) for helpful discussions 
and inspiration.


\begin{thebibliography}{99}
\bibitem{QL} The coefficients in such `quasi-linear' combinations are field-dependent, i.e., they are neither numbers, nor `group' constants.
\bibitem{XX} An obvious disadvantage of any Hamiltonian approach follows from the fact that all Hamiltonian-based methods are `frustratingly non-invariant',
since they essentially used the coordinate structure in the phase space (see, discussion in \cite{Arnold}).
\end{thebibliography}
\end{document}